\documentstyle[twoside,fleqn,espcrc2,graphics]{article}

\newcommand{\AmS}{{\protect\the\textfont2
  A\kern-.1667em\lower.5ex\hbox{M}\kern-.125emS}}

\newcommand{\ncom}{\newcommand}
\ncom{\bbbar}{{b\overline{b}}}
\ncom{\ccbar}{{c\overline{c}}}
\ncom{\MSbar}{{\overline{\rm MS}}}
\ncom{\Mbar}{{\overline{M}}}
\ncom{\Mbbar}{{\overline{M}_b}}
\ncom{\Mcbar}{{\overline{M}_c}}
\ncom{\Mk}{M_{\rm kin}}
\ncom{\Mx}{M_{\rm expt}}
\ncom{\nf}{n_{\rm f}}
\ncom{\Psub}{{\scriptscriptstyle  P}}
\ncom{\vect}[1]{{\bf #1}}
\ncom{\mvect}[1]{\mbox{\boldmath $#1$}}
\ncom{\be}{\begin{equation}}
\ncom{\ee}{\end{equation}}

\title{Bottom and Charm quark masses from lattice NRQCD}

\author{
NRQCD collaboration presented by
K.~Hornbostel%
\address{Dept. of Physics, Southern Methodist University, Dallas, TX 75275}%
\thanks{Work done in collaboration with 
C.~T.~H.~Davies, G.~P.~Lepage, C.~J.~Morningstar,
J.~Shigemitsu and J.~Sloan.  Support provided by
the DOE, NSF, PPARC, the Leverhulme Trust and the
Fulbright Commission.}
}
       
\begin{document}

\begin{abstract}
We present new values for the ${\rm \overline{MS}}$ 
masses of $b$ and $c$ quarks based on lattice NRQCD 
simulations of the $\Upsilon (\bbbar)$ and $\psi (\ccbar)$ 
systems.  These include three measurements of the $b$ mass 
based on quenched simulations with lattice spacings ranging 
from 0.05fm to 0.15fm, which we find to be largely independent 
of lattice spacing.  In addition, we find a consistent value 
from an unquenched simulation at 0.08fm.
\end{abstract}

\maketitle

\section{Simulations}

The $\Upsilon (\bbbar)$ and $\psi (\ccbar)$ systems possess several
properties which permit accurate lattice simulations.  They are physically 
small, allowing the use of small volumes.  They are insensitive to the 
presence of light quarks and to uncertainties in their masses.  They 
are well understood phenomenologically, which aids in the estimation of 
systematic errors.  Their spin-averaged mass splittings are insensitive to 
the $b$ or $c$ masses, allowing independent tuning of the lattice spacing
and bare masses.  Their properties have been precisely measured.
Finally, they may be efficiently simulated using a nonrelativistic 
effective action, NRQCD.  

To compute the spin-averaged spectra we employed an NRQCD action 
for the heavy quarks, improved to include leading and next-to-leading
order corrections for relativity and discretization errors.
We used the average plaquette to tadpole improve the link
operators.  Several collaborations generously provided gauge field
configurations for both zero and two light quark flavors, generated
using the standard (unimproved) Wilson action for gauge fields and staggered 
fermions for the light quarks.  We removed perturbatively the leading
discretization errors in the spectrum due to use of this unimproved 
gluonic action \cite{alpha}.
The details of our simulations and references for the gauge field
configurations appear in Ref.~\cite{simulations}.

\begin{figure}[htb]
\vspace{9pt}
\setlength{\unitlength}{0.240900pt}
\ifx\plotpoint\undefined\newsavebox{\plotpoint}\fi
\sbox{\plotpoint}{\rule[-0.200pt]{0.400pt}{0.400pt}}%
\begin{picture}(900,629)(0,0)
\font\gnuplot=cmr10 at 10pt
\gnuplot
\sbox{\plotpoint}{\rule[-0.200pt]{0.400pt}{0.400pt}}%
\put(220.0,113.0){\rule[-0.200pt]{0.400pt}{118.764pt}}
\put(220.0,113.0){\rule[-0.200pt]{4.818pt}{0.400pt}}
\put(198,113){\makebox(0,0)[r]{4}}
\put(816.0,113.0){\rule[-0.200pt]{4.818pt}{0.400pt}}
\put(220.0,212.0){\rule[-0.200pt]{4.818pt}{0.400pt}}
\put(198,212){\makebox(0,0)[r]{4.1}}
\put(816.0,212.0){\rule[-0.200pt]{4.818pt}{0.400pt}}
\put(220.0,310.0){\rule[-0.200pt]{4.818pt}{0.400pt}}
\put(198,310){\makebox(0,0)[r]{4.2}}
\put(816.0,310.0){\rule[-0.200pt]{4.818pt}{0.400pt}}
\put(220.0,409.0){\rule[-0.200pt]{4.818pt}{0.400pt}}
\put(198,409){\makebox(0,0)[r]{4.3}}
\put(816.0,409.0){\rule[-0.200pt]{4.818pt}{0.400pt}}
\put(220.0,507.0){\rule[-0.200pt]{4.818pt}{0.400pt}}
\put(198,507){\makebox(0,0)[r]{4.4}}
\put(816.0,507.0){\rule[-0.200pt]{4.818pt}{0.400pt}}
\put(220.0,606.0){\rule[-0.200pt]{4.818pt}{0.400pt}}
\put(198,606){\makebox(0,0)[r]{4.5}}
\put(816.0,606.0){\rule[-0.200pt]{4.818pt}{0.400pt}}
\put(220.0,113.0){\rule[-0.200pt]{0.400pt}{4.818pt}}
\put(220,68){\makebox(0,0){0}}
\put(220.0,586.0){\rule[-0.200pt]{0.400pt}{4.818pt}}
\put(357.0,113.0){\rule[-0.200pt]{0.400pt}{4.818pt}}
\put(357,68){\makebox(0,0){0.04}}
\put(357.0,586.0){\rule[-0.200pt]{0.400pt}{4.818pt}}
\put(494.0,113.0){\rule[-0.200pt]{0.400pt}{4.818pt}}
\put(494,68){\makebox(0,0){0.08}}
\put(494.0,586.0){\rule[-0.200pt]{0.400pt}{4.818pt}}
\put(631.0,113.0){\rule[-0.200pt]{0.400pt}{4.818pt}}
\put(631,68){\makebox(0,0){0.12}}
\put(631.0,586.0){\rule[-0.200pt]{0.400pt}{4.818pt}}
\put(768.0,113.0){\rule[-0.200pt]{0.400pt}{4.818pt}}
\put(768,68){\makebox(0,0){0.16}}
\put(768.0,586.0){\rule[-0.200pt]{0.400pt}{4.818pt}}
\put(220.0,113.0){\rule[-0.200pt]{148.394pt}{0.400pt}}
\put(836.0,113.0){\rule[-0.200pt]{0.400pt}{118.764pt}}
\put(220.0,606.0){\rule[-0.200pt]{148.394pt}{0.400pt}}
\put(45,359){\makebox(0,0){\rotatebox{90}{$\overline{M_b}(\overline{M_b})$ (GeV)}}}
\put(528,23){\makebox(0,0){$a$ (fm)}}
\put(220.0,113.0){\rule[-0.200pt]{0.400pt}{118.764pt}}
\put(706,541){\makebox(0,0)[r]{$n_{\rm f}=0$}}
\put(750,541){\raisebox{-.8pt}{\makebox(0,0){$\Diamond$}}}
\put(412,419){\raisebox{-.8pt}{\makebox(0,0){$\Diamond$}}}
\put(480,389){\raisebox{-.8pt}{\makebox(0,0){$\Diamond$}}}
\put(699,261){\raisebox{-.8pt}{\makebox(0,0){$\Diamond$}}}
\put(728.0,541.0){\rule[-0.200pt]{15.899pt}{0.400pt}}
\put(728.0,531.0){\rule[-0.200pt]{0.400pt}{4.818pt}}
\put(794.0,531.0){\rule[-0.200pt]{0.400pt}{4.818pt}}
\put(412.0,350.0){\rule[-0.200pt]{0.400pt}{33.244pt}}
\put(402.0,350.0){\rule[-0.200pt]{4.818pt}{0.400pt}}
\put(402.0,488.0){\rule[-0.200pt]{4.818pt}{0.400pt}}
\put(480.0,340.0){\rule[-0.200pt]{0.400pt}{23.608pt}}
\put(470.0,340.0){\rule[-0.200pt]{4.818pt}{0.400pt}}
\put(470.0,438.0){\rule[-0.200pt]{4.818pt}{0.400pt}}
\put(699.0,162.0){\rule[-0.200pt]{0.400pt}{47.698pt}}
\put(689.0,162.0){\rule[-0.200pt]{4.818pt}{0.400pt}}
\put(689.0,360.0){\rule[-0.200pt]{4.818pt}{0.400pt}}
\put(706,496){\makebox(0,0)[r]{$n_{\rm f}=2$}}
\put(750,496){\raisebox{-.8pt}{\makebox(0,0){$\Box$}}}
\put(497,369){\raisebox{-.8pt}{\makebox(0,0){$\Box$}}}
\put(728.0,496.0){\rule[-0.200pt]{15.899pt}{0.400pt}}
\put(728.0,486.0){\rule[-0.200pt]{0.400pt}{4.818pt}}
\put(794.0,486.0){\rule[-0.200pt]{0.400pt}{4.818pt}}
\put(497.0,300.0){\rule[-0.200pt]{0.400pt}{33.244pt}}
\put(487.0,300.0){\rule[-0.200pt]{4.818pt}{0.400pt}}
\put(487.0,438.0){\rule[-0.200pt]{4.818pt}{0.400pt}}
\end{picture}
\caption{$\MSbar$ $b$-quark mass $\Mbbar(\Mbbar)$ at various lattice 
     spacings $a$.  For $\nf=0$, $\beta = 6.2, 6.0$ and $5.7$, 
     respectively.  For $\nf=2$, $\beta = 5.6$.  }
\label{fig:Mb}
\end{figure}

\begin{table*}[hbt]
\caption{Perturbative parameters connecting the bare mass $M^0$ to 
         the $\MSbar$ mass $\Mbar(\mu)$.}
\label{tab:pertthy}
\begin{center}
\begin{tabular}{ccccccc}
 $\beta$ & $\nf$ & $aM_b^0$ &$a\mu$ & $c^{(1)}(\mu)$ & $aq^*$ & $\alpha_\Psub(q^*)$  \\
 \hline
    5.7 & 0   & 3.15   & 5 & -1.18   & 3.03   & .1911 \\ 
    6.0 &     & 1.71   &   & -1.63   & 5.73   & .1315 \\
    6.2 &     & 1.22   &   & -1.79   & 9.80   &  .1088 \\ \\
    5.6 & 2   & 1.80   &   & -1.61  & 5.34   &  .1570  \\ \\

         &    & $aM_c^0$ & \\
 \hline
    5.7  & 0  & 0.80   & 7 & -2.85   & 23.02  &  .1086 \\

\end{tabular}
\end{center}
\end{table*}

\begin{table*}[hbt]
\caption{$\MSbar$ masses for $b$ and $c$ quarks (preliminary). The lattice 
    spacing $a$ is determined from spin-averaged splittings between
    1P and 1S states.  Values for $M^0$ are from Eq.~\ref{m0}, rather
    than from $(aM^0)\,a^{(-1)}$, as discussed in the text.
    Errors are statistical and estimates of higher order relativistic 
    and discretization corrections, respectively.  Systematic errors are 
    combined in $M^0$ and $\Mbar(\Mbar)$.  The final error in $\Mbar(\Mbar)$ 
    estimates neglected higher-order perturbative contributions.  We do
    not include an error estimate to account for quenching.}
\label{tab:results}
\begin{tabular*}{\textwidth}{@{}l@{\extracolsep{\fill}}cccccc}
   $\beta$ & $\nf$ & $aM_b^0$ & $a^{(-1)}_{\rm 1P-1S}$~(GeV) & $a^{(-1)}_{\rm 2S-1S}$~(GeV) & 
    $M_b^0$~(GeV) & $\Mbbar(\Mbbar)$~(GeV) \\ 
 \hline
  5.7  & 0 & 3.15  & 1.41(4)(2)(5) & 1.36(13)(2)(4) & 4.22(5)(8) & 4.15(5)(8)(3)  \\ \\
  6.0  &   & 1.71  & 2.59(5)(3)(1) & 2.45(8)(3)(1) & 4.11(3)(2) & 4.28(3)(3)(3)  \\
       &   & 1.80  &               &               & 4.16(3)(2)   & \\
       &   & 2.00  &               &               & 4.21(3)(2)   & \\
       &   & 3.00  &               &               & 4.32(3)(2)   & \\ \\
  6.2  &   & 1.22  & 3.52(14)(4)(0)& 3.24(15)(4)(0) & 3.99(6)(2) & 4.31(6)(3)(3)  \\ \\
  5.6  & 2 & 1.80  & 2.44(6)(3)(1) & 2.38(10)(3)(1) & 4.08(4)(3) & 4.26(4)(3)(5) \\ 
\\

       &   & $aM_c^0$ &       &       & $M_c^0$     & $\Mcbar(\Mcbar)$ \\
 \hline
  5.7  & 0 & 0.80  & 1.23(4)(3)(8) & 1.20(20)(3)(8) & .98(3)(9) & 1.20(4)(11)(2)  \\ \\

 errors: &  &     & stat, rel, discr &  & stat, sys & stat, sys, pert \\ \\

\end{tabular*}
\end{table*}

\section{Extracting $\Mbbar$ and $\Mcbar$}

Results from these simulations allowed us to obtain a value 
for the $c$ quark $\MSbar$ mass $\Mcbar$ and values for the $b$ quark 
mass $\Mbbar$ at several lattice spacings, as follows.
We first measured the lattice spacing $a$ for each simulation by
comparing either the computed spin-averaged 1P-1S or 2S-1S splittings 
with data.  Because these splittings are very insensitive to the
bare mass $aM^0$, we were able to determine $a$ without first finely 
tuning the mass.  The results were accurate to within about 5\% for 
the $\Upsilon$ system, 15\% for $\psi$.

Using this $a$, we tuned the bare mass $a M^0$ so that the kinetic
mass $\Mk$, determined by fitting to the dispersion relation
\be
\label{disp}
 E(\mvect{p}) = E(0) + {\mvect{p}^2 \over 2 \Mk } - 
                         {\mvect{p}^4\over 8 \Mk^3} \, ,
\ee
agreed with experimental values for the $\Upsilon$ or $\eta_c$.

Once the fundamental parameters $aM^0$ and $a$ were fixed, we were able 
to convert quantities to physical units.  By using the expression
\be
\label{m0}
 M^0 = aM^0 \left({\Mx \over a\Mk}\right)
\ee
to express the bare mass in GeV, we minimized sensitivity
to uncertainties in $aM^0$ and $a$.  The bulk of $a\Mk$ is made up 
of $2 aM^0$, with dynamics providing the remainder.  As a result, 
the ratio ${aM^0 / a\Mk}$ is very insensitive to
errors in the dynamics from uncertainties in $aM^0$ and $a$; 
these represent only an error in an order $v^2$ correction to $1/2$.
As an extreme illustration, computing $M_b^0$ using a value for $aM_b^0$ 
as far off as $aM_c^0$ gives a result off by only 25\%.  In contrast,
using $a^{-1}$ directly in the form $M^0 = (aM^0) a^{-1}$ rather
than Eq.~\ref{m0} would unnecessarily promote errors in $a^{-1}$ by 
one power of the mass.

Refs.~\cite{gray} and \cite{colin} give perturbative
expressions relating the $\MSbar$ mass $\Mbar$ and the bare
mass $M^0$ to the pole mass, respectively.  We combined these 
to obtain 
\be
 \Mbar(\mu) = 
   M^0 \left[ 1 \; + \;  
   c^{(1)}(\mu) {\alpha_\Psub(q^*)\over \pi} \; +\;  \cdots \right] \; ,
\ee
with parameters listed in Table~\ref{tab:pertthy}.  
We used the scheme in Ref.~\cite{blm} to determine the
scale $q^*$ for $\alpha_\Psub$ in this relation.  In particular, 
we chose to extract $\Mbar(\mu)$ at a scale $\mu$ such that $q^*$ 
was well-defined. Values for $\alpha_\Psub$ came from measurements of 
the plaquette in these same simulations.

Finally, we ran $\Mbar(\mu)$ to its own scale using three-loop
evolution.  Our preliminary results are presented in 
Table~\ref{tab:results}.  
Insensitivity of the final result to the bare mass $aM_b^0$ is 
evident.  The systematic errors quoted account for higher-order 
relativistic and discretization corrections, most of which
were based on estimates from a potential model.  They do not account 
for errors associated with using the wrong number of light quarks.
Values for $\Mbbar$ at three lattice spacings and for zero and two 
light flavors $\nf$ also appear in Figure~\ref{fig:Mb}.

We also applied an alternate method to determine the $\Upsilon$ kinetic 
mass.  Rather than fitting to the dispersion relation above, 
we determined $a\Mk$ from
\be
 a\Mk = 2\, Z_m\, aM^0 + aE_{\rm NR} - 2\, aE_0 \, .
\ee
The first term on the right hand side gives twice the pole mass. 
The next two represent the energy for this state obtained in
the simulation with the self energy of the quarks subtracted,
which gives the binding energy.  The quark mass renormalization
$Z_m$ and self energy $aE_0$ were computed perturbatively to
one loop in Ref.~\cite{colin}.

We used this kinetic mass to convert the bare mass to physical 
units and obtained the $\MSbar$ masses as above.  Results
were consistent with the first method, but with larger errors due
to unknown higher-order terms in $Z_m$ and $aE_0$.

\section{Conclusions}

We presented new results for $\Mbbar$ and $\Mcbar$ from lattice
NRQCD calculations of the $\Upsilon$ and $\psi$ spin-averaged
spectra.  Our results for $\Mbbar$ showed little dependence on 
$a$ or on the presence of light quarks.  They are consistent with 
those presented at Lattice~97 and summarized in Ref.~\cite{lastyear}:
$\Mbbar = 4.15(5)(20)$ and $\Mcbar = 1.525(40)(125)$ from the APE 
group using HQET, and $\Mcbar = 1.33(8)$ from the FNAL group, using an
action which interpolates between relativistic and nonrelativistic 
regions.

\end{document}